\documentclass[10pt,journal,compsoc]{IEEEtran}
\ifCLASSOPTIONcompsoc
  \usepackage{cite}
\else
  \usepackage{cite}
\fi
\usepackage[perpage,symbol]{footmisc}
\usepackage{cite}
\usepackage[dvipsnames]{xcolor}
\usepackage{rotating}
\usepackage{mathtools}
\usepackage {amsmath}
\usepackage {amssymb}
\usepackage{algorithm,algpseudocode}
\usepackage {graphicx}
\usepackage {subcaption}
\usepackage {url}
\usepackage{multirow}
\usepackage{epstopdf}
\usepackage{mathtools}
\usepackage{diffcoeff}
\usepackage{mathtools}
\usepackage{gensymb}
\usepackage{fixltx2e}
\usepackage[T1]{fontenc}
\usepackage {amssymb,amsmath}
\usepackage{algorithm,algpseudocode}
\usepackage {graphicx}
\usepackage {url}
\usepackage{multirow}
\usepackage{multirow}
\usepackage{epstopdf}
\graphicspath {{../Figures}{../Graphs}}
\usepackage{mathtools}
\usepackage{amsmath}
\usepackage{gensymb}
\usepackage{breqn}
\usepackage{gensymb}
\usepackage[english]{babel}
\usepackage{microtype}

\begin{document}
\title{Attacking Deep Learning AI Hardware with Universal Adversarial Perturbation}

\author{Mehdi~Sadi,
        B. M. S. Bahar Talukder,
        Kaniz Mishty,
        and~Md Tauhidur Rahman
\IEEEcompsocitemizethanks{\IEEEcompsocthanksitem M. Sadi and K. Mishty are with the Department
of Electrical and Computer Engineering, Auburn University, Auburn, 36849, AL 36849, USA (e-mail: mehdi.sadi@auburn.edu)}
\IEEEcompsocitemizethanks{\IEEEcompsocthanksitem  B. M. S. Bahar Talukder and Md Tauhidur Rahman are with Florida International University, Miami, 33199, FL (e-mail: mdtrahma@fiu.edu) }}


\IEEEtitleabstractindextext{%
\begin{abstract}
Universal Adversarial Perturbations are image-agnostic and model-independent noise that when added with any image can mislead the trained Deep Convolutional Neural Networks into the wrong prediction.  Since these Universal Adversarial Perturbations can seriously jeopardize the security and integrity of practical Deep Learning applications, existing techniques use additional neural networks to detect the existence of these noises at the input image source. In this paper, we demonstrate an attack strategy that when activated by rogue means (e.g., malware, trojan) can bypass these existing countermeasures by augmenting the adversarial noise at the  AI hardware accelerator stage. We demonstrate the accelerator-level universal adversarial noise attack on several deep Learning models using co-simulation of the software kernel of Conv2D function and the Verilog RTL model of the hardware under the FuseSoC environment.
\end{abstract}

}

\maketitle
\IEEEdisplaynontitleabstractindextext
\IEEEpeerreviewmaketitle

\section{Introduction}
\label{sec:Intro}

\IEEEPARstart{T}{he} last decade has been a conspicuous success in the history of AI. Using Deep Neural Network (DNN), AI has superseded human intelligence in many of the applications: such as Deepmind's AlphaGo has defeated the world's best human Go player, AlphaFold can successfully predict the 3D protein fold structure, which had been a challenge for past 50 years  \cite{AlphaFold}, since 2015 ImageNet classifier models have been outperforming human in object recognition  \cite{sze2017efficient}. The rapid advancement in model building and their promising accuracy in particular tasks have led DNN models to be frequently deployed in a wide range of applications, including many safety-critical areas such as biometric security, autonomous vehicle, cyber-security, health, and financial planning. To withstand the overarching AI-based applications, which require massive computations, the demand for specially optimized hardware for these extremely power-hungry and data-driven computations has also surged both in the data centers (for training) and edge devices (for inference)  \cite{jouppi2017datacenter,intl_vpu}. As these AI accelerator devices are being used in several safety-critical applications, it is essential to ensure their security and integrity. 

While AI and Deep Learning becomes pervasive in all aspects of our life, we should not disregard the fact that the DNN models are highly vulnerable to adversarial examples \cite{szegedy2014intriguing}. An adversarial example is an example that is intentionally created to be misclassified by an DNN model by adding adversarial perturbation to the input data \cite{szegedy2014intriguing}. The addition of this adversarial perturbation to clean inputs makes the input falsely classified by state-of-the-art DNN models while retaining their credibility to the human. Adversarial samples are obtained by adding imperceptibly small perturbations to a correctly classified input image so that it is no longer classified correctly. The adversarial samples expose fundamental blind spots in the DNN training algorithms. The adversarial examples represent low-probability (high-dimensional) “pockets” in the input space, which are hard to efficiently find by simply randomly sampling the input around a given example \cite{szegedy2014intriguing, goodfellow2015explaining}. By using a simple optimization procedure adversarial examples can be found. The \emph{generalization} (i.e., adversarial perturbation created for a model trained with one data distribution is also harmful to the models which are trained over different sets of data distribution) and \emph{transferability} (i.e., the adversarial examples computed for one model are also valid for every other model with completely different architectures, hyperparameters, and training data) property of adversarial examples have doubly contributed to this stealthy nature of DNN models   \cite{szegedy2014intriguing, goodfellow2015explaining, tramer2020ensemble, Moosavi_Dezfooli_2017_CVPR}.

The existence of the above described feature of DNN gives enough opportunity to an attacker to intrude an AI model to inflict massive damage in his intended domain. Depending on the extent of information (such as model parameter, input, output) available to the adversary and in which phase the attack has been launched, various types of attacks emerged in the last several years. In a poisoning attack, a malicious model is developed during the training phase either by training the model on adversarial data distribution, altering the training procedure, or manipulating the model's weights or architecture  \cite{Truong_2020_CVPR_Workshops,gu2017badnets}. In contrast, in evasion attack, an adversary discovers malicious inputs on which the model will make unexpected error \cite{Truong_2020_CVPR_Workshops}. When the adversary has access to all sorts of necessary information about the model such as model architecture, model parameters, and weights, the attack scenario is termed as white-box attack  \cite{PBAA2017}. Contrarily, in black-box attack, an attacker doesn't have access to any model information other than model's response to chosen input \cite{PBAA2017}. Upon its discovery, it has been shown that the adversarial examples can be exploited to successfully attack all kinds of application of deep learning \cite{yuan2019adversarial}. However, all of these attacks are software-based and mainly focus on devising innovative techniques to generate adversarial examples. It is assumed that the hardware, where the DNN models are computed on (training and inference), is always reliable and trustworthy. Unfortunately, this assumption is no longer valid in the current semiconductor industry and security  vulnerabilities in the hardware, as well as software malware, calls for serious attention from the research community  \cite{guin2014counterfeit,contreras2013secure}. 

The AI/Deep Learning hardware can be compromised from both the hardware (e.g., Trojans) and software (e.g., malware) based methods. Because of ASICs' complex and intricate structure, the semiconductor industries involve design out-sourcing and fabrication globalization. In most cases, the designed GDSIIs travel overseas, from designer to foundry, before they are finally deployed in the target device. This whole flow of the supply chain has become one of the most prominent causes of the infiltration of untrusted hardware in the semiconductor industry \cite{guin2014counterfeit}. Anyone with an evil intent engaged in this flow is capable of any malicious alteration of the target product. Because of this reason hardware security has received attention in the past decade, however very few works have been done in the AI specialized hardware platform. From the software perspective, malicious code in the form of malware  \cite{Malware_1, Malware_2, Malware_3} can compromise Deep Learning/AI computing systems. Although malware protection techniques and computer system security have significantly improved over the past decade, the attackers are consistently coming up with new evasive and sophisticated strategies to breach the software and hardware level malware detection techniques to execute malicious functions  \cite{Malware_1, Malware_2, Malware_3}. Preventing the attackers from installing malicious programs for gaining privileged access to the system is practically impossible as they can exploit various techniques such as web browser vulnerabilities,  social engineering (e.g., phishing), email attachments, flash drives, etc. Symantec reported that 246,002,762 new malware variants emerged in 2018  \cite{Malware_1}. Adversarial noise can be injected into Deep Learning AI hardware by compromising its operating system and execution software with various malware attacks similar to Stuxnet  \cite{Stuxnet}.

Precisely in our attack strategy, we exploit the recently discovered phenomena that there exists a \textit{Universal Adversarial Perturbation (UAP)} noise which when added to the inputs of the DNN model can severely compromise their prediction accuracy  \cite{Moosavi_Dezfooli_2017_CVPR,Liu_2019_ICCV}. The following scenario can be pictured as an example. An inference-only device is used for the inference tasks of a pre-trained AI model in an autonomous vehicle. The task of the model is to provide an appropriate decision about surrounding objects and road signs to the vehicle. The DNN model is authentic. However, the device which is performing the computation has a severe security flaw- it has a \textit{Universal Adversarial Noise} image stored in its memory which can be augmented to the input image captured by the vehicle’s camera and is capable of fooling almost all DNN models irrespective of their architecture and training data distribution. The model will then decide on the perturbed image and provide the disastrous decision to the car based on its misclassified output.

The existing Fault-Injection and Bit-flip-based attacks  \cite{clements2018hardware, rakin2019bit, laserfault,liu2020imperceptible,zhao2019fault} require a fair amount of knowledge about the networks' architectures and parameters to materialize  the attack by careful manipulation of the network. These constraints make these attacks hard to implement in real-time. In contrast, the \textit{UAP} attacks are more generic, and hence more destructive. In this paper, we present a novel accelerator-based DNN attack that does not need any information about the model's architecture and parameters. The attacker only needs to have the scope of planting malware or software/hardware Trojans.

The key contributions and highlights of this paper are:
\begin{itemize} 
\item To the best of our knowledge, this work for the first time proposes and demonstrates an accelerator-based DNN attack which requires little-to-no knowledge about the target DNN model by exploiting \textit{Universal Adversarial Perturbation (UAP)}. The proposed attack is sneaky enough to obscure its existence yet powerful enough to cause massive damage.

\item We propose a novel technique to interleave the UAP noise with the actual image to mislead the trained DNN model when the attack is activated with malware or software/hardware Trojans. Since our technique avoids the usual methods of adversarial noise injection at the sensor or camera level and directly injects at the accelerator hardware stage, it is more stealthy.

\item We provide a detailed comparative analysis on the complexity, stealth, and implementation challenges of the existing trojan, fault-injection and bit-flip based, and the proposed UAP based attacks on the AI/Deep Learning hardware.
\end{itemize}

The rest of the paper is organized as follows. Section 2 presents the background on DNNs, accelerators, and adversarial noise attacks. Section 3 presents the Threat model of the UAP attack. Section 4 presents the details of the UAP interleaving method. The stealth of the proposed UAP attack is discussed in Section 5. Experimental results are presented in Section 6, followed by related work in Section 7.

\section{Background}
\subsection{AI/Deep Learning Neural Networks}
\begin{figure}[h]
\centering
\includegraphics[width=3.6in,height=3in]{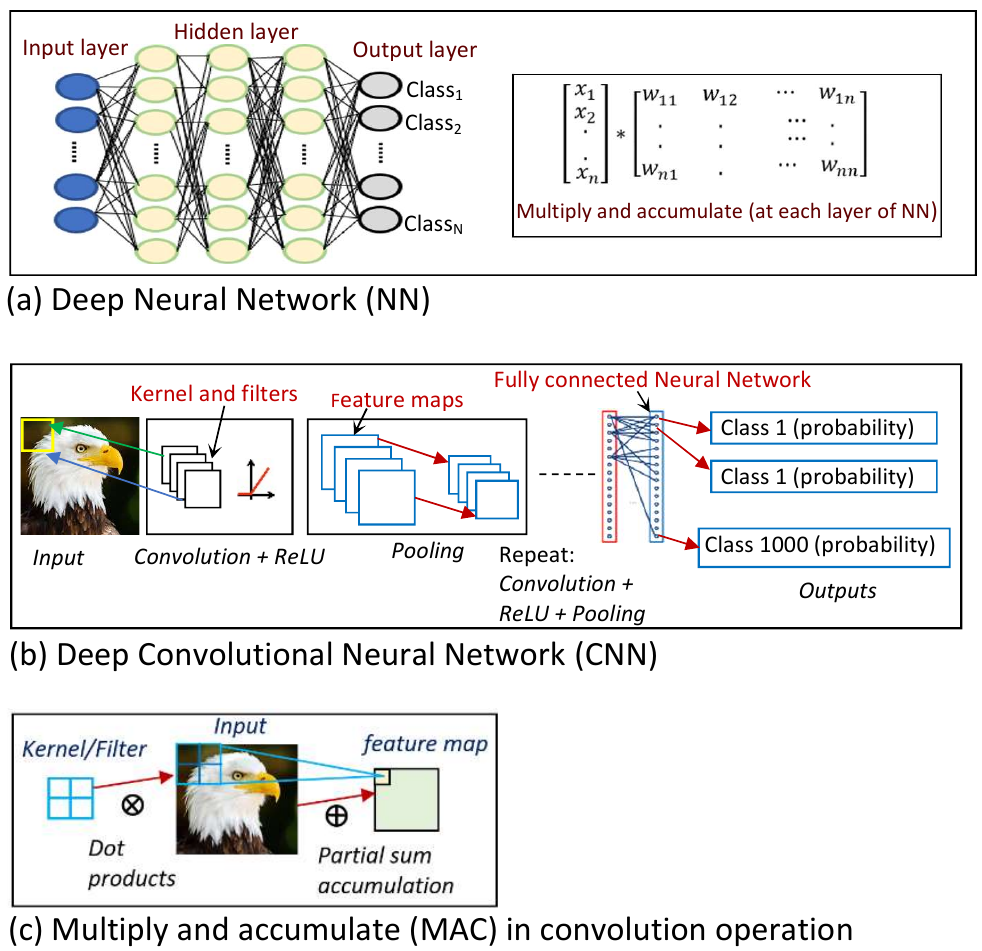}
\caption {(a) Deep NN; (b) Deep CNN; (c) Convolution in CNN.}
\label{fig:cn}
\vspace{-0.05in}
\end{figure}

At the essence of AI/Deep Learning algorithms are the backpropagation-based NN and CNN. As shown in Fig. \ref{fig:cn} (a), a deep NN consists of an input layer, followed by several hidden layers and a final output layer. Depending on the data size, the complexity of training, dropout, and pruning rate, some layers in the NN are fully connected and others sparsely connected \cite{sze2017efficient}. The connection strengths between the adjacent layers are represented by a weight matrix $W$, and the matrix parameters $w_i$  are learned by the backpropagation-based learning equation, $\Delta w_i=-\alpha*\frac{\partial Error}{\partial w_i}$, where $\alpha$ is learning rate and $Error$ is the prediction error. During the forward pass of training and inference phases, the output activation of a layer, $X_o$, is obtained by multiplying the input activation vector with the weight matrix followed by addition of a bias term, and finally passing the result through an non-linear function such as ReLU,  $X_o=ReLu(W*X_i + b)$. 

Due to their higher accuracy, deep CNNs have become the standard for image and pattern recognition \cite{sze2017efficient}. The operation of a deep CNN is briefly shown in Fig. \ref{fig:cn}(b). During training and inference, each image or pattern is convolved successively with a set of filters where each filter has a set of kernels. After ReLU activation and pooling, the convolution operation is repeated with a new set of filters. Finally, before the output stage, fully connected NNs are used. The convolution operation is shown in Fig. \ref{fig:cn}(c) and it consists of dot products between the input feature-maps and filter weights ($h$), mathematically, $f_{out}(m,n)=\sum_{j}\sum_{k} h(j,k)f_{in}(m-j,n-k)$.

\vspace{-1ex}

\subsection{AI/Deep Learning Accelerator Architecture}
\label{sec:Arch}
\vspace{-2ex}
\begin{figure}[h]
\centering
\includegraphics[width=3.5in,height=1.9in]{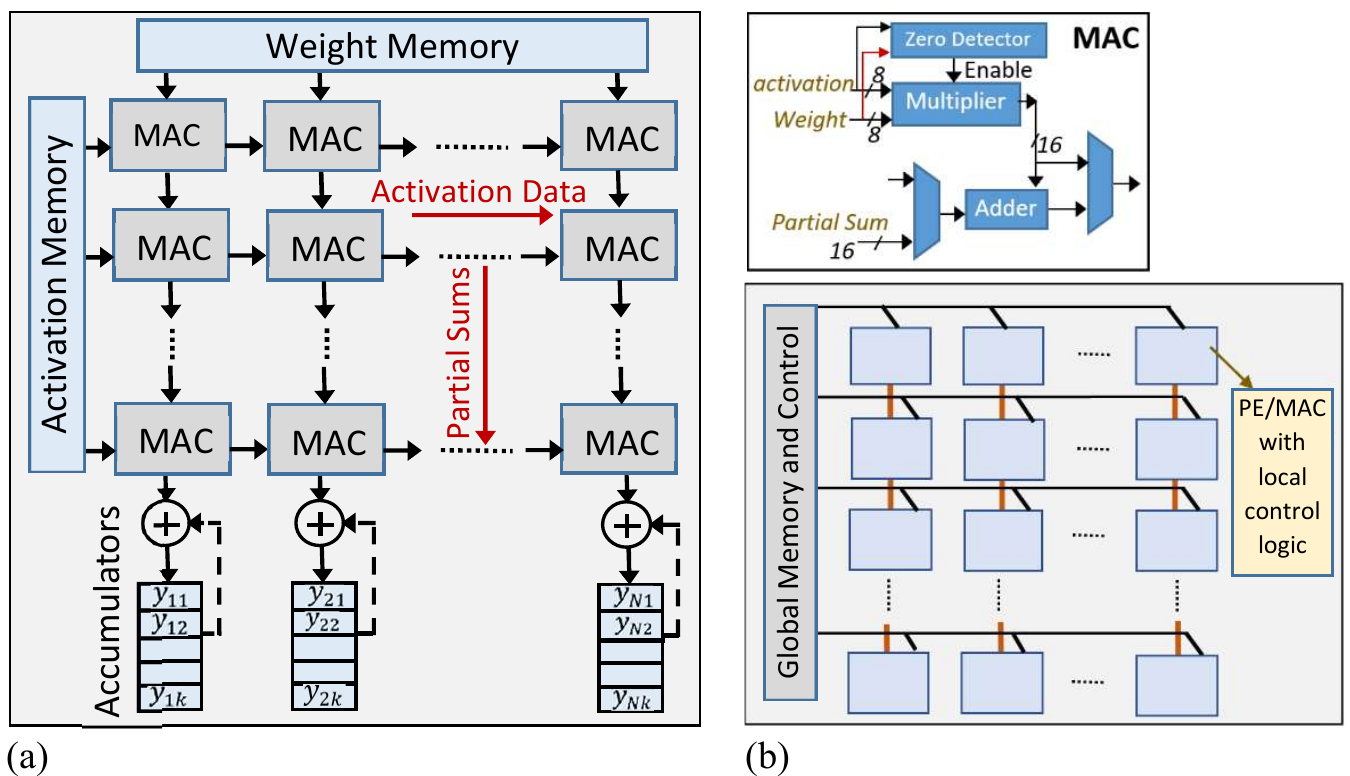}
\caption {AI accelerator with PE/MAC arrays. (a) Systolic architecure; (b) SIMD architecture.}
\label{fig:PE}
\end{figure}

Since the computations in deep NN/CNN are mostly dominated by Multiply and Accumulate (MAC) operations, the AI accelerators are primarily occupied with arrays of Processing Elements (PE) optimized for fast MAC function \cite{sze2017efficient}. As shown in Fig. \ref{fig:PE}, the accelerator architectures can be categorized into two domains,  (i) tightly-coupled 2D systolic-arrays (e.g., Google’s TPU) \cite{sze2017efficient,jouppi2017datacenter}, and (ii) loosely coupled spatial arrays with independent PEs interconnected with NoC/mesh and using  SIMD architecture \cite{sze2017efficient}. 

\subsection{Adversarial Perturbation to Mislead Trained AI Models} 
The well-trained state-of-the-art deep learning models can be fooled to misclassify inputs by augmenting the input images with adversarial patterns. Adversarial samples are created by adding imperceptibly small and non-random perturbations to the correctly predicted original input images to cause the trained deep learning models to falsely-classify the input images \cite{szegedy2014intriguing,goodfellow2015explaining,tramer2020ensemble,Moosavi_Dezfooli_2017_CVPR}.  Because of the small magnitude of the perturbation, the adversarial samples are often imperceptible to the human eye and correctly classified by a human observer, but fails at the trained deep learning model.  The adversarial samples are distinct from randomly distorted samples. In the majority of the cases, the randomly distorted image samples - where the magnitude of the random noise is not too large compared to the actual image -  are classified correctly by the state-of-the-art deep learning networks, however, the adversarial samples are almost always misclassified \cite{szegedy2014intriguing,goodfellow2015explaining,tramer2020ensemble,Moosavi_Dezfooli_2017_CVPR}. Two important properties of adversarial perturbations are as follows \cite{goodfellow2015explaining}.

\noindent
\textit{- Generalization Property:} Deep learning models demonstrate consistency in misclassifying adversarial samples.  Adversarial images created for one deep learning model is often erroneously classified by other models with different hyperparameters (i.e., different model architecture, number of layers, initial weights, regularization, etc.), and  a different subset of training data \cite{szegedy2014intriguing,goodfellow2015explaining,tramer2020ensemble,Moosavi_Dezfooli_2017_CVPR}. 

\noindent
\textit{- Transferability Property:} The adversarial samples are robust and demonstrate transferability property \cite{szegedy2014intriguing, goodfellow2015explaining, tramer2020ensemble,Moosavi_Dezfooli_2017_CVPR}. Adversarial samples crafted to mislead a specific deep learning model are also effective in misclassifying other deep learning models even if their architectures greatly differ.

These two properties make adversarial attacks more pervasive and stealth. The adversarial patterns can be maliciously augmented with the input data at the AI hardware level, and cause the trained deep learning model to fail.

\subsubsection{Adversarial Sample Crafting}   There are two classes of methods for generating adversarial samples. 

\noindent
\textit{- Image-based Adversarial Perturbation:} In this method \cite{szegedy2014intriguing, goodfellow2015explaining, tramer2020ensemble}, adversarial samples are created on a per-image basis  by adding carefully-crafted noise to the original image along the gradient directions. Once a particular image is augmented with this noise it can show its adversarial efficacy across different neural network models. However, to achieve a successful adversarial goal, the perturbation needs to be generated separately for each image \cite{szegedy2014intriguing, goodfellow2015explaining}.

Using the Fast Gradient Sign Method (FGSM),  adversarial samples can be created in this approach \cite{ goodfellow2015explaining, tramer2020ensemble}. The adversarial input $\overrightarrow{x^*}$ can be generated from the input pattern $\overrightarrow{x}$ by adding a perturbation, i.e., $\overrightarrow{x^*}=\overrightarrow{x}+\overrightarrow{\eta}$, The perturbation is obtained with FGSM using the equation,  
$\overrightarrow{\eta} = \epsilon*sign(\nabla_{\overrightarrow{x}} J(\theta, \overrightarrow{x},y))$
Where $\theta$ is model parameter, $\overrightarrow{x}$ is input to the model, $y$ is the target and $J(\theta,x,y)$ is the cost function. The gradient is calculated using backpropagation \cite{szegedy2014intriguing, goodfellow2015explaining}.

\noindent
\textit{- Universal Adversarial Perturbations:}  In the Second method \cite{Moosavi_Dezfooli_2017_CVPR}, universal adversarial perturbations are generated based on the input (i.e., image) data distribution rather than individual images. In addition to being network-agnostic (i.e., transferable among different state-of-the-art deep learning models), these universal perturbations are image-agnostic and maintain their adversarial efficacy across different images belonging to the same distribution (e.g., ImageNet data set).
For a classification function $f$ that outputs  a predicted label $f(x)$ for each image $x \in \mathbb{R}^d$, a perturbation vector $v \in \mathbb{R}^d$ is the universal adversarial perturbation vector that misleads the classifier $f$ on almost all datapoints sampled from the distribution of images in $\mathbb{R}^d$ such that $f(x+v) \neq f(x)$ for most $x$ in the distribution \cite{Moosavi_Dezfooli_2017_CVPR}. The algorithm to find $v$  proceeds iteratively over a set of images sampled from the distribution and gradually builds the universal perturbations. Due to their small magnitude, the perturbations are hard to detect and do not significantly alter the data distributions. The universal perturbations were shown to have generalization and transferability property across different architectures on ImageNet data set. This implies that to fool a new image on an unknown deep neural network a simple augmentation of a universal perturbation generated on AlexNet/VGG16 architecture is highly likely to misclassify the image.

\subsection{Adversarial Attack Strategy during Deep Learning Inference} 
The adversarial attacks on trained deep learning models can be broadly classified as white-box and black-box types. In the white-box attack scenario, for a target deep learning model, the adversary has access to all the elements of the training procedure such as the model architecture, training data, training algorithm, weights, etc. With this information, in a white-box attack scenario, the adversary can use arbitrary methods to create the adversarial samples \cite{tramer2020ensemble}. In a stronger black-box attack scenario, the adversary only has access to the deep learning model's input-output relationships. In the black-box attack in \cite{PBAA2016} it was demonstrated that adversarial attacks can be manifested by developing a surrogate model of the target DNN under attack by knowing only the output class information of the model for various inputs (i.e., Oracle query \cite{tramer2020ensemble}). In this black-box attack mode - without truly knowing the architecture of the model being attacked - using the surrogate model the attacker can generate adversarial samples. The transferability property  (i.e., adversarial sample created on one deep learning model is also effective on another) of adversarial perturbation ensures that the samples will be effective on the actual DNN under attack \cite{tramer2020ensemble}.

For the case of Universal Adversarial Perturbation based attacks \cite{Moosavi_Dezfooli_2017_CVPR}, the attack strategy gets even simpler. Because of the image-agnostic universal nature of the perturbation, by only knowing that the AI/Deep Learning model is doing image recognition/classification tasks, the universal adversarial perturbations can be created by the attacker. For example, universal adversarial perturbations created based on the large ImageNet data set demonstrates its adversarial effectiveness across various state-of-the-art image classification deep learning models by virtue of the Generalization and Transferability properties of adversarial samples \cite{Moosavi_Dezfooli_2017_CVPR}.   
  
To counteract adversarial attacks of the Deep Learning models, adversarial training was proposed where adversarial samples were used during the training procedure. However, in \cite{tramer2020ensemble} it was demonstrated that the adversarilly trained models remained vulnerable to multistep and elaborate black-box attacks where perturbations created on undefended models were transferred to the adversarilly trained models.

\section{Threat Model}
\noindent
\textbf{Attacker Knowledge:} The attack model is BlackBox \cite{PBAA2016} as it is not required to know the model parameters, training data, etc. The attacker only needs to know the Deep Learning model`s task category (e.g., image classifier) and use the appropriate universal adversarial perturbation. In our threat model to materialize the attack it is not required for the trained model to have back-doors \cite{PBAA2016, Strip, SentiNet}. These make the attack more stealth.

\vspace{0.1in}
\noindent
\textbf{Adversarial Goal:} Once the attack mode is activated with malicious means (e.g., malware) the adversarial noise is injected into the AI hardware, and the attacker gains the ability to fool the pre-trained AI/Deep Learning model into misclassifying the input samples during inference.

\begin{table}[]
\caption{ Strategy for attacking AI hardware accelerator with universal adversarial perturbation}
\begin{tabular}{|l|l|}
\hline
\textbf{Attack Type}                                                                                       & \textbf{Attack Strategy (Details in Section IV)}                                                                                                                                                                                                                                                                        \\ \hline
\begin{tabular}[c]{@{}l@{}}Malicious Noise\\ Interleaving  and \\ Convolution (MNIC)\end{tabular} & \begin{tabular}[c]{@{}l@{}}Universal adversarial noise is interleaved \\ with the original image and the filter rows \\ are repeated before the first convolution \\ operation. Malicious modification \\ (e.g., with malware) of the  inputs of\\   ‘Conv2D’ function can accomplish this task.\end{tabular} \\ \hline
                                                                                                                    
\end{tabular}
\end{table}

\vspace{0.1in}
\noindent
\textbf{Attack Strategy:} The strategy for attacking AI hardware accelerator with universal adversarial perturbation is shown in Table 1. To implement this attack in the AI accelerator hardware, with heightened privilege (root access on Linux, or admin access in Windows), the adversary can access and modify the target program`s address space through the \textit{/proc/[PID]/map} and through .DLL (Dynamic Link Library)  injection in Unix and Windows operating systems, respectively \cite{Forensics_1, Forensics_2, LiveTrojan}. Specific types of malware such as Trojan, rootkit, and fileless can covertly alter system library to accomplish malicious program execution \cite{Malware_1, Malware_4}. Trojan type malware such as Emotet uses functionality that helps evade detection by some anti-malware products \cite{Malware_1}. Fileless malware (e.g., the Astaroth malware) are unique as they are capable of altering files related to the operating system without installing any files of their own \cite{Malware_4}. As the targets of the Fileless malware are part of the operating system, they often appear as legitimate files to the antivirus/malware tools, and this makes the attacks stealthy \cite{Malware_4}. One of the most dangerous malware, Rootkit \cite{rootkit}, gives privileged access of the computing system to the attacker. Once the rootkit is installed in the system the adversary can maintain privileged access and full control over the system without detection as it can disable malware detectors. A rootkit can target specific applications and partially modify their execution behavior by injecting its own code. Another approach to malicious code/function injection is to remap memory between processes with a malicious kernel module, which has proved effective in many well-known attacks like Stuxnet \cite{Stuxnet}.

\section{Accelerator-level Addition of Adversarial Perturbation }

\begin{figure}[h]
\centering
\vspace{-0.05in}
\includegraphics[width=3.5in,height=1.2in]{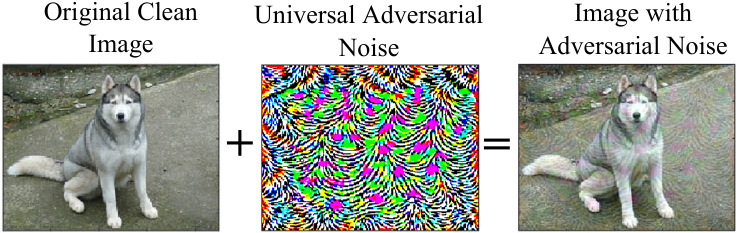}
\vspace{-0.05in}
\caption {Universal adversarial perturbation added with clean image to deceive the trained deep learning model.}
\label{fig:adv_pert}
\end{figure}

\begin{figure}[h]
\centering
\includegraphics[scale=1]{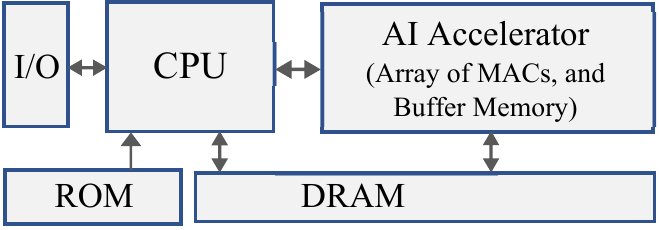}
\caption {AI hardware accelerator architecture.}
\label{fig:Fig_hardware}
\end{figure}

To fool the trained AI/Deep Learning models the adversarial perturbation noise needs to be added with the input image before it is processed with convolution filters. In a regular setting to superimpose the adversarial patterns on the image,  modifications at the camera or the input image source itself is required before the AI accelerator processes it. Since this input alteration adversarial attack is well-known, the image capture sensors and digitization modules are protected with security measures to prevent any adversarial noise additions. In our hardware-level attack, we assume the image sources are protected and clean, and in our attack model the universal adversarial perturbations are added to the clean images by the attacker at the SoC level by exploiting the internal memory system and software with malware/Trojan. A simplified block diagram of an AI accelerator hardware system is shown in Fig. \ref{fig:Fig_hardware}. The host CPU receives the clean input images through the input port or sensors and then sends those to the accelerator hardware through the DRAM for efficient and fast processing. In our attack model, the universal adversarial pattern is stored on the SoC at the ROM or some other on-chip non-volatile memory. The adversarial pattern can also be transferred to the system externally as malware. Under malicious attack, this adversarial pattern is added with each input image before it is processed at the AI accelerator hardware. Because of the transferability and generalization properties of universal adversarial perturbations (i.e., discussed in Section IIC), a single well-crafted pattern can fool many images with high-efficiency \cite{Moosavi_Dezfooli_2017_CVPR}.

\begin{figure}[h]
\centering
\includegraphics[width=3.6in,height=2.6in]{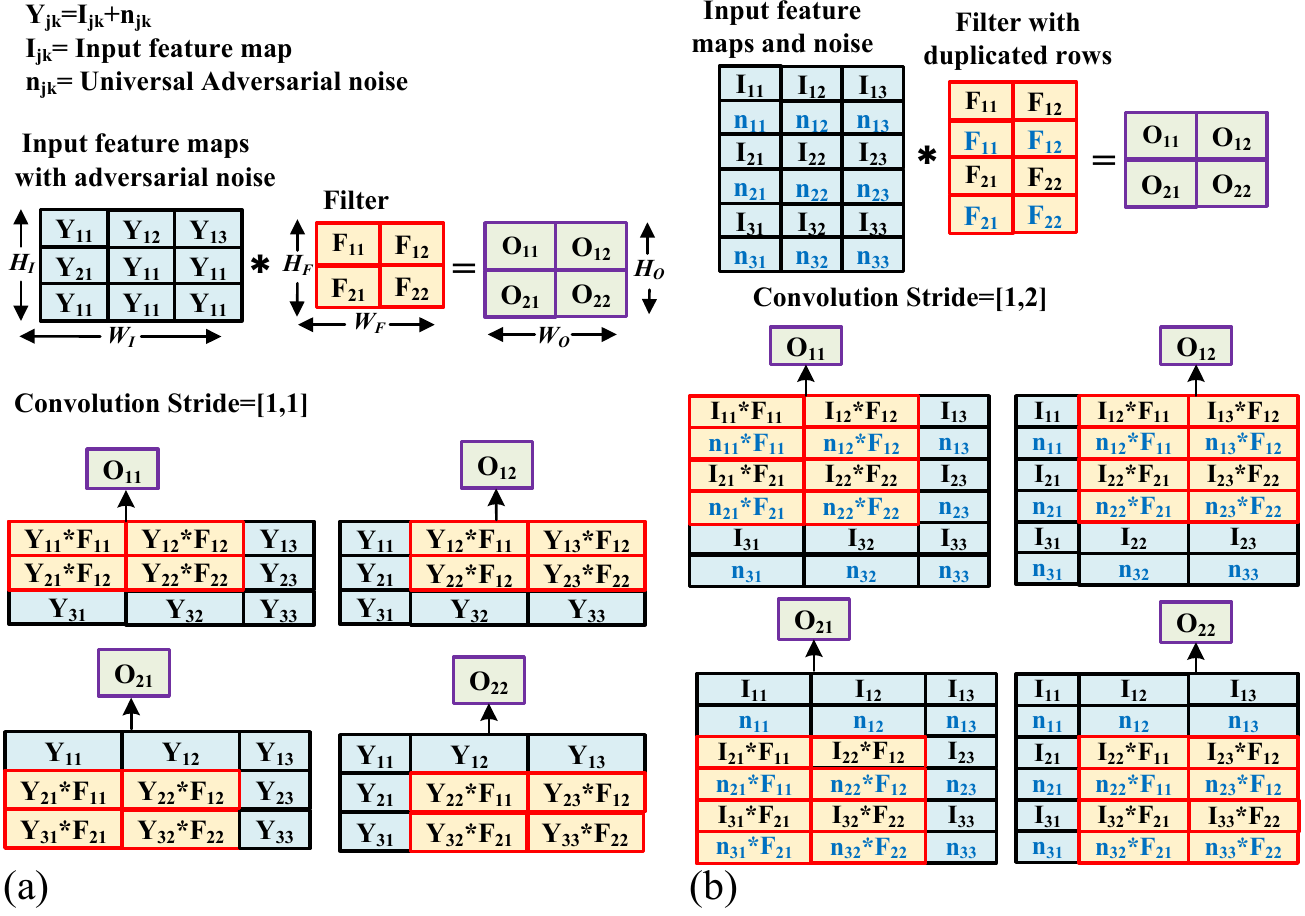}
\caption {Using the additive property of convolution, the same output feature map can be generated by noise interleaving without directly adding noise with the input. (a) Convolution with direct adversarial noise addition with input; (b) Noise interleaved with input, filter duplicated, and vertical stride doubled to accomplish the same task.}
\label{fig:Fig_Noise_Conv}
\end{figure}

\begin{figure}[h]
\centering
\includegraphics[scale=1]{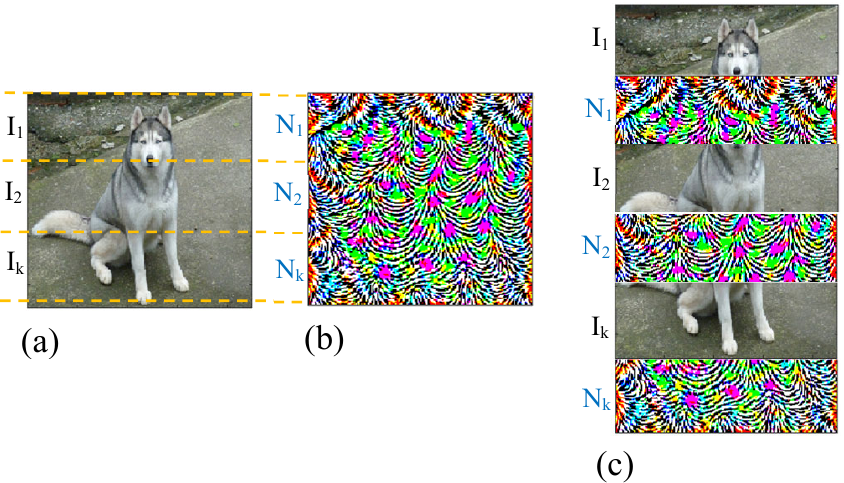}
\caption {(a) Clean input image  ; (b) Universal adversarial perturbation (c) Adversarial perturbation interleaved with the input image.}
\label{fig:Fig_adv_noise_added}
\end{figure}

Directly adding the adversarial noise with the input image using the CPU`s ALU will require many additions which may raise the security alarm indicating a malicious attack. For example, for typical RGB images of 3 channels (e.g., from ImageNet data set) $224 \times 224 \times 3$ addition operations will be required at the ALU to add the adversarial noise with the input before sending it to the AI accelerator. As a result, in our attack strategy, we avoid directly adding the stored universal adversarial perturbation with the image at the host CPU. We utilize the additive property (i.e., $f*(a+b)=f*a+f*b$, where $*$ is the Convolution operator) of convolution operation to create the same effect as convolving the filter with the image perturbed with adversarial noise, without directly adding the noise to the image. This operation is mathematically explained with the illustration of Fig. \ref{fig:Fig_Noise_Conv}. In Fig. \ref{fig:Fig_Noise_Conv}(a), $Y_{jk}$ is the noise added image pixels, i.e., $Y_{jk}=I_{jk} + n_{jk}$, and $I_{jk}$ and $n_{jk}$ are clean image and noise respectively for pixel location $jk$. The output of convolution with filter $F$ is $O$. In  Fig. \ref{fig:Fig_Noise_Conv}(b), instead of directly adding the noise pixel $n_{jk}$ with corresponding clean image pixels $I_{jk}$, the noise pixel rows are interleaved with the image pixels. To produce the exact same outputs of Fig. \ref{fig:Fig_Noise_Conv}(b), $O_{mn}$, two other adjustments are necessary in Fig. \ref{fig:Fig_Noise_Conv}(b), (i) duplication of filter rows, (ii) doubling the vertical stride of convolution as shown in Fig. \ref{fig:Fig_Noise_Conv}(b). With this modification, we can produce the same effect of convolving the filter with noise added inputs, without ever explicitly adding the noise. In Fig. \ref{fig:Fig_adv_noise_added}, an example is shown where universal adversarial noise patterns are interleaved across rows with an image from the ImageNet data set.

We need to perform this (i) noise interleaved and (ii) repeated filter row-based convolution only for the first layer of the deep learning model, and the rest of the layers follow the regular approach of convolution. Since we are doubling the filter rows in this approach,  the number of MAC operations will also double only for the first layer of the model. However, in contrast to the ALU of the CPU, the AI accelerator is equipped with a large number of MAC modules (e.g., Google TPU has 65K MACs \cite{jouppi2017datacenter}) and hence the extra MAC operations will complete very fast without raising any security alarm. Moreover, because of zero-skip (i.e., if any of the MAC inputs is zero, the operation is skipped), and the presence of a variable number of zeros in digitized images and quantized filter weights, the number of MAC operations are not deterministic (i.e., fixed) for clean images. As a result of the presence of a large number of MAC arrays in AI accelerators and zero-skipping in modern accelerators,  the extra MAC operations (i.e., only in the first layer) of our noise interleaved convolution will get masked to the extent not to raise any security alarm.

\begin{figure}[h]
\centering
\includegraphics[scale=0.95]{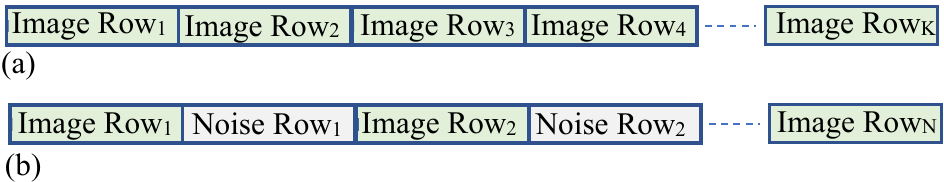}
\caption {Input image data positioning in Memory, (a) regular scenario  ; (b) under adversarial attack noise data is interleaved.}
\label{fig:Fig_Interleav}
\end{figure}

The AI accelerator loads the input image in its global buffer memory from the DRAM \cite{jouppi2017datacenter, sze2017efficient}. In the regular scenario, the input image rows are placed adjacently in the Memory as shown in Fig. \ref{fig:Fig_Interleav}(a). In our proposed hardware-level attack scheme, the rows of universal adversarial perturbation are placed adjacent to the image rows in the Memory as shown in Fig. \ref{fig:Fig_Interleav}(b). This can be achieved by malware or software/hardware Trojans.  As a result, during execution at the accelerator the noise data are interleaved with the clean image data and ensuing convolution in the accelerator in effect convolves the filter with images corrupted with universal adversarial perturbation as graphically shown in Fig. \ref{fig:Fig_Noise_Conv}(b).

The hardware-level malicious addition of the adversarial perturbation attack can be initiated with malware or software-controlled trojans \cite{MTrojan, HIT, softhat}. Additionally, the attacker can utilize the unused instruction in the instruction set to manifest the adversarial noise injection attack.

\section{Stealth Against Adversarial Detection Techniques}
Our attack strategy can evade conventional protections against adversarial attacks in AI/Deep Learning systems as discussed in detail in the following. 

\subsection{SGX Independence} Intel SGX \cite{sgx} can only protect the data and code segments stored in the \textit{Enclave}. However, in our threat model,  (1) the adversarial noise data (about 74KB), and (2) malicious "Conv2D" function (few lines of code) are generic and already known to the attacker. As a result, the attacker adds these data and code segments to the original version even before they are transferred to the protected \textit{Enclave}.   In our threat model, the attacker does not need to modify data in \textit{Enclave}, thus the attack scenario is independent of SGX.

\subsection{Row-hammer Independence} In the conventional Bit-flip attack (i.e., where the attacker needs to modify the trained model's weights) \cite{rakin2019bit}, Row-hammer-based weight bit flipping in DRAM is necessary to materialize the attack. With Row-hammer protected DRAM circuit design and Intel SGX, the DRAM-based Bit-flip attacks can be thwarted. Our threat model is non-invasive to the DNN model as we do not need to tamper with the original trained model parameters.

\subsection{Bypassing Conventional Countermeasures against Adversarial Noise} In conventional attack models, the adversarial noise is generally added at the image source, hence protective measures are taken accordingly. However, in our attack strategy adversarial noise is added internally at the AI accelerator level. As a result, it can bypass the state-of-the-art adversarial detection techniques that assume adversarial noise is added at the image source, making the attack stealthier. The SentiNet method presented in \cite{SentiNet} detects adversarial attacks such as Trojan trigger, backdoor, poisoning, and adversarial patches on test images. SentiNet first identifies the salient regions of an image that highly influences its class labels, and therefore are potential targets of adversarial noise addition. To identify if the test image is compromised, these salient regions are superimposed on benign golden samples and tested if they are misclassified upon overlaying. High misclassification rates indicate the presence of adversarial noise or patches. The STRIP technique proposed in \cite{Strip} detects trojaned images on the concept that if an image is backdoored with adversarial patches it will always classify into a certain adversary-chosen class. The STRIP detection algorithm adds perturbations to the test image and identifies if its class label changes, and a static class label indicates that the image was trojaned with an adversarial patch. However, to implement these approaches at run-time additional AI accelerator hardware is needed to execute these detection models in parallel to the original deep learning model, and this incurs extra energy and latency. Moreover, these detection techniques are developed for scenarios where the adversarial noise or trojan is localized to a patch, in contrast to our threat model where low magnitude universal noise spans across the image. 

In \cite{Akhtar_2018_CVPR} a defense mechanism against UAP has been presented. First, a Perturbation Rectifying Network (PRN) is trained and then used as the ‘pre-processing’ layer of the targeted model. Next, a perturbation detector is separately trained as a binary classifier to extract discriminative features from the difference between the inputs and outputs of the PRN. However,  this proposed method requires extra pre-trained models to be added before the target model, and also assumes a conventional scenario where the images are compromised with UAP before they are processed at the accelerator. As in our proposed attack the UAP is interleaved with the images at the hardware accelerator stage, the detection techniques of \cite{Akhtar_2018_CVPR} can be evaded and the attack becomes stealth.

\subsection{Non-deterministic Execution}
In our approach because of duplicating the filter rows of the first layer, the number of MAC operations in the first layer of the model can double. But this may not raise any security alarm, because in modern AI accelerators zero-skipping is implemented \cite{sze2017efficient} where if any of the ifmap or filter input is zero, MAC operation is skipped.  Because of Image digitization and quantization before feeding to the accelerator, there are several zeros in the input image pattern and this number varies from image to image. Moreover, there are also zeros present in the filter. As a result, the number of MAC operations per image is not deterministic, and hence the extra MAC required in our adversarial attack will vary from image to image without showing any constant pattern that can raise alarm.

\section{Experimental Results }
To generate the universal adversarial perturbations, we used the tools available from \cite{Moosavi_Dezfooli_2017_CVPR}. The normalized maximum magnitude of the adversarial perturbation was kept within 5\% of the normalized maximum magnitude of the images. For our analysis, we used the standard 50K validation image samples from ImageNet database. After adding the universal adversarial perturbations with these images, we used the pre-trained version of several widely used Deep Learning models and analyzed the fooling rate of adversarial noise with PyTorch \cite{pyt}. In accordance with \cite{Moosavi_Dezfooli_2017_CVPR} we define the fooling rate as the percentage of image samples that altered their labels under adversarial noise compared to clean images (i.e., irrespective of ground truth). We used the same (i.e., a single pattern of 224 x 224 pixels with 3 channels) universal adversarial noise (i.e., transferability property as discussed in Section II) for our experiments with AlexNet, ResNet-50, VGG-16, and GoogleNet deep learning models. The fooling rates are shown in Table \ref{Tab:foolrate}.  We can observe that more than 80\% of the validation image samples altered their labels when adversarial noise was added.

\begin{table}[]
\centering
\caption {Adversarial Fooling Rate}
\begin{tabular}{|c|c|c|c|c|}
\hline
Network                                                                  & AlexNet & VGG-16 & ResNet-50 & GoogleNet \\ \hline
\begin{tabular}[c]{@{}c@{}}Adversarial \\ Fooling Rate (\%)\end{tabular} & 90.8    & 88.9   & 84.2      & 85.3      \\ \hline
\end{tabular}
\label{Tab:foolrate}
\end{table}

\begin{figure}[h]
	\centering
	\includegraphics[scale=0.95]{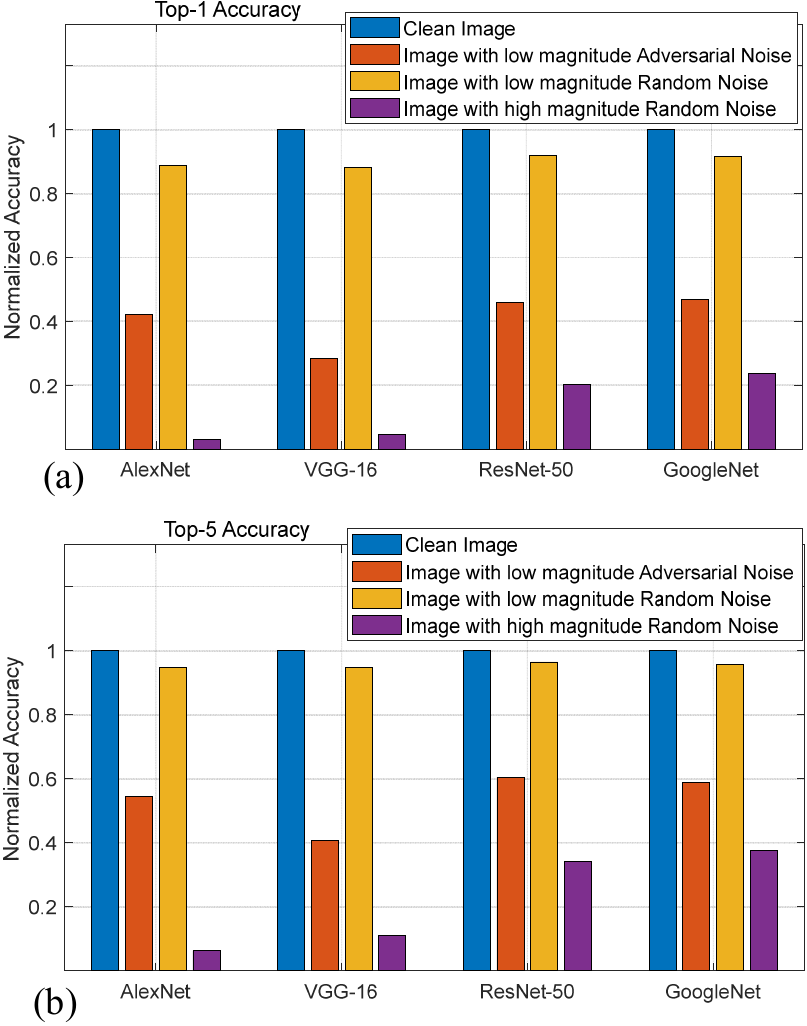}
	\caption {Normalized prediction accuracy changes for (i) clean, (ii) adversarially perturbed, (iii) low magnitude random noise augmented, and (iv) high magnitude random noise added images for ImageNet benchmark. (a) Top-1 accuracy;  (b) Top-5 accuracy. }
	\label{fig:Fig_Top1_5_Accuracy}
\end{figure}

The normalized Top-1 and Top-5 accuracy changes for ImageNet data under the universal adversarial perturbation are shown in Fig. \ref {fig:Fig_Top1_5_Accuracy}. Additionally, to demonstrate the efficacy of universal adversarial noise compared to random noise, we also ran  two sets of additional experiments where,  (i) the clean images were augmented with low magnitude random noise (i.e., noise magnitude within 5\% of the original image, similar to the case of the adversarial perturbation). (ii) high magnitude (i.e., maximum allowed noise magnitude same as the image magnitude) random noise added with the clean images. Results for AlexNet, VGG-16, ResNet-50 and GoogleNet, are shown in Fig. \ref{fig:Fig_Top1_5_Accuracy}. From these experiments, it can be seen in Fig. \ref{fig:Fig_Top1_5_Accuracy} that random noise with an amplitude similar to adversarial perturbation only slightly impacts the accuracy, implying that the sate-of-the-art AI/deep learning models are highly immune to small random noise at the input. However, carefully crafted universal adversarial noise of the same low magnitude range (i.e., 5\% of the original image)  can drastically compromise the fidelity and accuracy of the models. To achieve similar accuracy degradation effects the random noise magnitude must be large (i.e., on the same order as the image) as shown in Fig. \ref{fig:Fig_Top1_5_Accuracy}. However, compared to the low magnitude adversarial noise, augmenting high magnitude random noise at the AI accelerator level is a complex task because of the larger bit-width and memory size requirements (i.e., 4 bits per pixel for adversarial noise versus 8 bits for random noise), and this raises practical implementation challenges from a hardware-based attack perspective. 

With a large magnitude noise, the amount of non-zero bits in the digitized pattern will be significant and comparable to the image itself, as a result, hardware accelerator level malicious addition of the noise pattern can raise security alarm. On the contrary, universal adversarial patterns can induce significant accuracy degradation, however, because of their much lower magnitude compared to the actual image they can be injected more secretively. For example for images from the imageNet dataset with 224x224x3 pixels and 8 bits per pixel, to achieve the same level of image misclassification as the adversarial noise with high magnitude (e.g., noise magnitude atleast 50\% of image magnitude) random noise,  90\% more noise bits are required.

Moreover, because of the lower magnitude of the adversarial noise, the pattern mostly consists of zeros and the low number of non-zero bits only occur in the lower-level bits (e.g., LSBs) of the digitized adversarial pattern. For ImageNet benchmark images with 8 bits per digitied pixel of the clean image, the pixels of the low magnitude (i.e., noise magnitude limited within 5\% of the original image magnitude)  universal adversarial perturbation can be digitized with maximum 4 bits. Also, the low bit-width adversarial noise can be compressed using conventional sparse weight compression techniques of quantized AI models \cite{sze2017efficient}, and easily hidden in the AI hardware for later malicious deployment in a stealthy manner.


\begin{figure}[h]
	\centering
	\includegraphics[width=3.6in,height=2.2in]{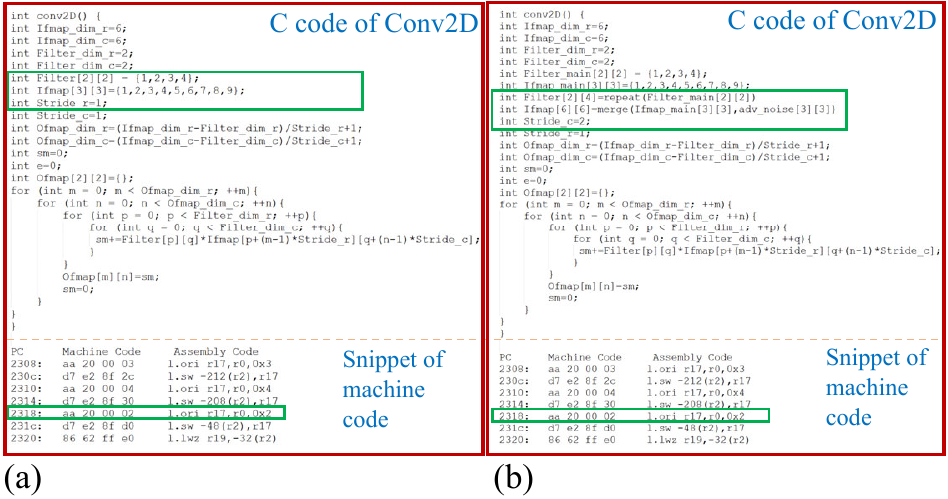}
	\caption {An example of Conv2D code. Under attack, the machine code can be compromised to inject universal adversarial perturbation in the images. (a) Regular code.  (b) Merging adversarial perturbation by image interleaving and altering the stride.}
	\label{fig:Fig_C_Code}
\end{figure}

To demonstrate the hardware-level adversarial attack on AI/Deep Learning models based on the interleaved augmentation of universal adversarial perturbation with images (i.e., described in Section III)  on a SoC platform, we used the OpenRISC based MOR1K soft CPU core with FuseSoC platform \cite{fusesoc}. FuseSoC platform allows writing high-level C,  and assembly code  programs which can be compiled and executed on the MOR1K CPU with other added hardware modules. First, we present the C programming code example of Conv2D function (i.e., the convolution function used in Convolutional Neural Networks) to demonstrate how input image and adversarial noise data interleaving, filter row duplication and convolution stride modifications can be accomplished. In Fig. \ref{fig:Fig_C_Code}(a), the C code for a regular Conv2D task with $Stride=1$ is shown. Filter row duplication and interleaving of the adversarial perturbation with image before feeding it to the Conv2D task is shown in the C code of Fig. \ref{fig:Fig_C_Code}(b).  After running the MOR1K toolchain \cite{fusesoc} under the FuseSoC environment, the C code is converted to equivalent machine code as shown in the snippets in the bottom segments of Fig. \ref{fig:Fig_C_Code}, where it can be seen that how the $Stride$ parameter can be altered in the machine code. 

The attack can be manifested both at the high-level C code kernel stage and at the compiled machine code stage. Based on the illustrative example of \ref{fig:Fig_C_Code}, to implement the attack strategy in C code we can add $if-else$ command-based malicious $jump$ instructions to move the program counter to malicious code segments where the input image data is interleaved with the adversarial noise,  filter rows duplicated and $Stride$ doubled before performing convolution operation. Similarly, the attack can be accomplished by direct code injection/alteration in the compiled machine code. We assume some external malware  or trojan \cite{Stuxnet,Malware_1} will initiate this malicious code alteration to activate the injection of adversarial perturbations. In our experiment, we executed the software-level malware/trojan attack by developing a custom script that altered (i.e., with root access \cite{rootkit}) the machine code after compilation and before execution in the hardware. Please note that the details of the software-level code injection mechanisms are outside the scope of this paper. Since multiple prior research in software security has demonstrated such code alteration and injection attacks \cite{Stuxnet,Malware_1,Malware_4,rootkit}, we assume such malware attacks to modify code are also practical in our context.
 
The universal adversarial perturbation requires only a single sample that has dimensions similar to the input image (e.g., 224 X 224 X 3) but of much lower bit size. This common perturbation sample can be stored in the SoC’s ROM or other non-volatile memory by the adversary. Once the adversarial attack is activated at the hardware-level by the attacker, the noise pattern bits are interleaved with the input image bits and stored in the DRAM. From DRAM, this noise-interleaved image is supplied to the AI accelerator core. In our experiment with FuseSoC, we used the RTL implementation of a Systolic array of Multiply and Accumulate (MAC) units as the accelerator. We modeled the memory holding the adversarial noise patterns with Verilog, the accelerator is written in Verilog RTL, and the Conv2D kernels were written in C code. Verilog PLI routines were used for cycle-accurate simulation. We used the Synopsys VCS tool along with the FuseSoC flow.

\section{Related Work and Comparison} 
In this section, we briefly review the existing methods of compromising the security and integrity of Deep Learning/AI models.

\subsection{Trojan Based Attacks}
Trojan and fault-injection are powerful tools for adversarial attacks \cite{torres2017fault}. In a Trojan attack, an attacker aims to discover ways to change the behavior of the model in some circumstances so that existing behavior remains unchanged. Under the Trojan attack, the system works perfectly unless the Trojan is being triggered \cite{hu2020practical}. The Trojan backdoor can be embedded in either the DNN model (e.g., the parameters), the hardware platform (e.g., CPU, GPU, accelerator), or the software stack (e.g., the framework) \cite{hu2020practical}. Existing Trojan attacks are mainly algorithm-based or system-based \cite{liu2018sin,ji2019programmable}. In the algorithm-based Trojan attacks, attackers leverage the intrinsic vulnerability of the DNN model and intercept the training set in order to train the model in a specific structure or weight parameters so that the DNN can be crashed when the triggering options (for example, input with specific markers or patterns) activate the Trojan \cite{liu2018sin,ji2019programmable}. In the system-based Trojan attack, attackers exploit hardware and software-stack to hide and activate Trojan \cite{liu2018sin,li2018hu}.  Trojans inserted during training can force the AI model to deviate into the adversarial mode.

\subsection{Fault Injection Attack}
Because of the distributed and redundancy structure of DNNs, they are usually robust against noisy inputs, and therefore it is usually assumed that they are tolerant against fault attacks \cite{torres2017fault}. However, these faults can occur in all major parameters of a DNN, such as weights, biases, inputs, etc., and can significantly degrade NN accuracy. Liu et al. \cite{liu2017fault} introduced a single bias attack to modify one parameter in DNN for misclassification through fault injection. In their proposed work, they observed that some parameters linearly control the DNN outputs. Venceslai et al. Zhao et al. \cite{zhao2019fault} proposed a fault sneaking attack to modify model parameters and mislead the DNN. The effectiveness of laser fault in DNN adversarial attack was demonstrated in \cite{laserfault}. They found that laser fault injection attack on DNN misclassification is as effective as other attacks. Liu et al. \cite{liu2020imperceptible} injected glitch to fool the DNN by perturbing the clock signal to the PE array infrequently. This attach is stealthy because it does not leave any sign of attack. However, the attacker must have the capability to perturb the clock signals.

\subsection{Bit-flip Attack}
In bit-flip attack \cite{rakin2019bit}, attackers deterministically induce bit flips in model weights to compromise DNN inference accuracy by exploiting the rowhammer vulnerability \cite{kim2020revisiting}. The bit-flip-based \textit{un-targeted} attack misclassifies all inputs into a random class \cite{rakin2019bit}. On the other hand, the bit-flip-based \textit{targeted} attacks on weight parameters mislead selected inputs to a target output class. However, for both \textit{un-targeted} or \textit{targeted} attacks, the attacker needs the knowledge of DNN architecture. If the attacker does not know the DNN architecture and other parameters, the bit-flip-based attack is ineffective \cite{rakin2019bit}.

Rowhammer is a circuit-level Dynamic Random-access Memory (DRAM) vulnerability used to cause unauthorized memory changes. In the rowhammer attack methodology, an attacker injects faults in the target memory cells through repeated access to the same row (aggressor row) before the target DRAM row gets refreshed.  Lately, rowhammer has also been investigated in adversarial AI attacks \cite{rakin2019bit,liu2017fault}. Although a bit-flip-based attack from rowhammer shows promise, an actual attack in the real world is far behind from a proof-of-concept exploit performed in a research lab \cite{cojocar2020we}. First, not all DRAM chips are vulnerable to rowhammer attack \cite{kim2014flipping}. Second, a successful rowhammer attack requires exhaustive offline preparation (for example, knowing the exact physical layout \cite{kim2020revisiting} that varies from manufacturer to manufacturer and model to model even if they are from the same manufacturer). Third, the structure of physically adjacent rows impacts the rowhammer attack. For example, we need to hammer more aggressor rows if a victim row follows a half-row pattern than a victim row that is contiguous within a single physical row \cite{cojocar2020we}. Fourth, there has been numerous hardware-, and software-based approaches to defend a system from rowhammer attack \cite{cojocar2020we}. DRAM memory manufacturers also claim that their DRAM chips/modules are robust against rowhammer attacks \cite{cojocar2020we}. These memory chips are mostly equipped with error-correcting code (ECC), higher refresh rate, target row refresh (TRR), or partial TRR to make their DRAM robust against rowhammer attack \cite{noauthor_mitigations_2015}.  Fifth, recent studies suggest that most of the flips happen from \textit{one} to \textit{zero}, and only a few of them flip from \textit{zero} to \textit{one} \cite{cojocar2020we}. Therefore, a designer can distribute the weights in DRAM cells in such a way that the flips from rowhammer will have minimal impact on DNN classification accuracy. Sixth, spin-transfer torque magnetoresistive RAM (STT-MRAM) that has the potential to replace DRAM is not shown vulnerable to rowhammer yet. Therefore, the bit-flip-based adversarial attack is not possible for STT-MRAM-based DNN accelerators \cite{stt_acc}.

\subsection{Comparison of Universal Adversarial Perturbation Attack with other attacks}
\noindent
\textbf{Black-box vs. White-box Attack:} The fault-injection and Bit-flip-attack (BFA), demands full access to the deep learning model’s weights and gradients. Thus, these are considered as a white-box attack. In contrast, the universal perturbation-based attacks are Black-box type, and only need to know what type of data the deep learning model is analyzing \cite{Moosavi_Dezfooli_2017_CVPR}. Due to the transferability property, a single adversarial pattern can be effective across multiple deep neural networks that are classifying images.

\noindent
\textbf{Accuracy:} In terms of accuracy degradation, since the BFA attack is tailored for each specific deep learning model (e.g., white-box) and follows a defined weight bit-flip sequence, the accuracy degradation of it is much severe compared to universal adversarial perturbation-based attacks. 

\noindent
\textbf{Complexity of Hardware Implementation:} Practical implementation of BFA or fault-injection on the AI hardware`s memory system is extremely complex due to the nature of the deterministic sequence of bit patterns needed to be flipped \cite{rakin2019bit,liu2017fault}. The Row-hammer-based DRAM bit-flip requires that the attacker is aware of the exact location of the target weight (e.g., byte offset in DRAM) bits in DRAM and they can be hammered independently. Moreover, the weight bits must reside in DRAM long enough such that the adversary can perform the Row-hammer attack. However, with increasing on-chip buffer memory capacity in modern AI accelerators/GPUs, the weight residence time in DRAM is very short. Especially, for modern pruned and AI models, the weights are sparse and compressed, as a result exactly identifying and confining each weight within DRAM pages is very complex. In \cite{rakin2019bit,liu2017fault}, the authors show that flipping non-target bits with fault injection or rowhammer bit-flip can not impact the accuracy of DNNs. In contrast, our hardware-level universal adversarial perturbation attack attack method is much simpler to implement. Our attack model only requires that malware or Trojan maliciously manipulate the Conv2D code for the first layer of the deep learning model, and the rest of the deep learning layers run as usual.

\noindent
\textbf{Detectability:} Practical Row-hammer-based implementation of BFA is very challenging and Row-hammer must span over multiple bits over many layers of the Deep Learning model. The extensive perquisite DRAM memory manipulation, and consecutive access to the same memory address in Row-hammer can raise security alarm. In contrast, our proposed hardware-level implementation of adversarial perturbation attack is much simpler and relatively stealth. 
\vspace{-2ex}
\section{Conclusions}
\label{sec:conc}
With the pervasive application of AI/Deep Learning, AI hardware systems are becoming mainstream. Adversarial attacks are a popular and effective method to compromise the fidelity of Deep Learning/AI models. In this paper, we demonstrate an AI hardware-level attack that can secretly add universal adversarial perturbation to clean input images and fool the Deep Learning model to misclassify the data. The adversary can initiate the attack with malware or Trojan to interleave a common adversarial perturbation (stored on-chip) with input samples before processing at the AI accelerator hardware. By simple malicious alteration of the program counter (e.g., jump to a malicious code segment of Conv2D), the attack can be manifested without raising security alarms. We demonstrate the attack using state-of-the-art deep learning models, and OpenRISC based FuseSoC platform with Systolic array-based accelerator hardware.

\ifCLASSOPTIONcompsoc
\else
\fi

\bibliographystyle{IEEEtran}
\tiny
\bibliography{References.bib}

\begin{IEEEbiographynophoto}{Mehdi Sadi}
(S'12-M'17) is an Assistant Professor at the Department of Electrical and Computer Engineering (ECE) at Auburn University, Auburn, AL.  Dr. Sadi  earned his PhD in ECE from  University of Florida, Gainesville, USA in 2017, MS from University of California at Riverside, USA in 2011 and BS from Bangladesh University of Engineering and Technology in 2010.   Prior to joining Auburn University, he was a Senior R\&D SoC Design Engineer in the Xeon Design team at Intel Corporation in Oregon. Dr. Sadi`s research focus is on developing algorithms and Computer-Aided-Design (CAD) techniques for implementation, design, test \& reliability of AI, and brain-inspired computing hardware. His research also spans into developing Machine Learning/AI enabled System-on-Chip (SoC) design flows, and Design-for-Reliability for safety-critical AI hardware systems. He has published more than 20 peer-reviewed research papers. He was the recipient of Semiconductor Research Corporation best in session award and Intel Xeon Design Group recognition awards.
\end{IEEEbiographynophoto}

\begin{IEEEbiographynophoto}{B. M. S. Bahar Talukder}
(S'18) obtained his Ph.D.  in Electrical and Computer Engineering at Florida International University in 2021. He received his Bachelor's degree from Bangladesh University of Engineering and Technology, Dhaka, Bangladesh. His primary research interests include hardware security and reliability, secured computer architecture, machine-learning application in system security, and emerging memory technologies. He joined Apple as senior engineer. 
\end{IEEEbiographynophoto}

\begin{IEEEbiographynophoto}{Kaniz Mishty}
received the B.S. degree in Electronics and Communication Engineering from Khulna University of Engineering and Technology, Bangladesh, in 2018. She is currently working towards her Ph.D. degree in ECE at Auburn University, AL, USA. Her current research interests are energy and area efficient VLSI system design, AI/Neuromorphic hardware design and AI/ML in CAD. As a summer intern, she worked on incorporating AI/Machine Learning in ASIC design flows at Qualcomm, Santa Clara.
\end{IEEEbiographynophoto}

\begin{IEEEbiographynophoto}{Tauhidur Rahman} (S'12–M'18) is an Assistant Professor with the Electrical and Computer Engineering Department, the Florida International University. He
received a Ph.D. degree from the University of
Florida in 2017, and the master’s degree from
the University of Connecticut in 2015. His
current research interests include hardware security and trust, machine learning, embedded security, and reliability. He received the
NSF CRII Award, in 2019.
\end{IEEEbiographynophoto}

\end{document}